\documentstyle[amssymb,aps]{revtex}

\begin{document}
\draft
\title{Confined Harmonically Interacting Spin-Polarized Fermions in a Magnetic
Field: Thermodynamics}
\author{S. Foulon, F. Brosens, J.\ T. Devreese\thanks{%
Also at Universiteit Antwerpen (RUCA), and at Technische Universiteit
Eindhoven, NL 5600 MB Eindhoven, The Netherlands.}}
\address{Departement Natuurkunde, Universiteit Antwerpen (UIA), Universiteitsplein 1,%
\\
B-2610 Antwerpen}
\author{L. F. Lemmens}
\address{Departement Natuurkunde, Universiteit Antwerpen (RUCA), Groenenborgerlaan\\
171, B-2020 Antwerpen}
\date{Submitted 13 October 1998, resubmitted 22 December 1998 to Phys. Rev. E}
\maketitle

\begin{abstract}
We investigate the combined influence of a magnetic field and a harmonic
interparticle interaction on the thermodynamic properties of a finite number
of spin polarized fermions in a confiment potential. This study is an
extension using our path integral approach of symmetrized density matrices
for identical particles. The thermodynamical properties are calculated for a
three dimensional model of $N$ harmonically interacting spin polarized
fermions in a parabolic potential well in the presence of a magnetic field.
The free energy and the internal energy are obtained for a limited number of
particles. Deviations from the thermodynamical limit become negligible for
about 100 or more particles, but even for a smaller number of fermions
present in the well, scaling relations similar to those of the continuum
approximation to the density of states are already satisfied.
\end{abstract}


\section{Introduction}

In the present paper we study the thermodynamical properties of a confined
system of spin-polarized fermions in the presence of a magnetic field. The
method used is an extension of the combination of the path integral
formalism \cite{Feynman1} and the method of symmetrized density matrices 
\cite{Feynman2}, developed previously \cite{PRE1,PRE2,PRE3,PRE4} for a model
system of harmonically interacting identical particles (bosons or fermions)
in a parabolic well (hereafter for brevity referred to as the harmonic
model).

Because of the experimental realization of Bose-Einstein condensation \cite
{Anderson,Davis,Bradlet} and the theoretical work on this phenomenon
employing other methods \cite
{Grossman1,Grossman2,Ketterle,Kirsten,Haugerud,Cohen,Krauth,Minguzzi}, full
details for this harmonic model with interaction were first worked out for
bosons. The model shows the onset of Bose-Einstein condensation in the
specific heat \cite{SOLCOM2} for a finite number of particles, and its
moment of inertia is drastically reduced below the condensation temperature 
\cite{PRA2}. An application of the method to real systems can be found in
Ref. \cite{SOLCOM3} for $^{87}Rb$.

The actual calculations for the fermion case require more advanced
techniques, such as the generating function approach and the corresponding
contour integration, because of a numerical sign problem. In the absence of
a magnetic field, explicit results for the thermodynamics and the static
correlation functions of the harmonic model of spin-polarized fermions were
already obtained with these techniques \cite{PRE3,PRE4}.

The harmonic model clearly has intrinsic value on its own, because it is one
of the rare examples of an exactly soluble many-particle system with
interaction. The physics of the model is relatively straightforward in the
sense that it allows for center of mass excitations that oscillate at
frequencies different from those of the internal degrees of freedom. This
property makes it well suited as a trial model for the variational treatment
of the thermodynamics of systems with more realistic interactions because
the model parameters can be related with the system characteristics with the
aid of the Jensen-Feynman inequality \cite{Feynman2}. The present paper
adresses only the first part of such an approach because it requires also
the density and the pair correlation function which we could obtain under
the simplifying assumptions of no magnetic field \cite{PRE4}. Furthermore,
it provides a testing ground for new approaches to Monte Carlo simulations
of interacting fermions such as many body diffusion \cite{PRE5,PLA1,SOLCOM1}%
. Especially for quantum dots, it is important to take the magnetic field
into account in order to freeze out the opposite spin states. In the present
paper we present an extension of the methods mentioned above to harmonically
interacting confined fermions in a magnetic field.

The paper is organized as follows. In Sec. II we present the path integral
for harmonically interacting particles in a parabolic confinement potential
in the presence of a homogeneous magnetic field. This will be done for
distinguishable as well as for identical particles. The mathematical details
of the calculation for identical particles are given in Appendix A. In Sec.
III the permutation symmetry will be taken into account with the aid of the
projection technique. The introduction of the permutation symmetry implies
the rewriting of the sum over all possible permutations to a cyclic
summation \cite{Feynman2} which leads to the generating function of the
partition function. Specific results for fermions will be presented in Sec.
IV. This involves the extraction of the partition function and other
thermodynamical quantities from the generating function. Also the ground
state energy and the magnetic susceptibility in the zero-temperature limit
will be investigated. Additionally we will study the finite number
corrections to the thermodynamic limit for the free energy and the internal
energy as a function of temperature and magnetic field. In the last section
some conclusions are given.

\section{Identical oscillators in a magnetic field}

The calculation of the path integral for $N$ identical interacting
oscillators in a magnetic field is similar to the case without a magnetic
field in \cite{PRE1}. This approach crucially relies on the detailed
investigation of the classical action, and the path-integral corrections to
this classical action. Alternatively, a more stochastic approach could be
followed \cite{Simon}. The Lagrangian (in atomic units) for $N$ oscillators
with harmonic two-body interactions and in the presence of a homogeneous
magnetic field is given by 
\begin{equation}
L=\frac{1}{2}\sum_{j=1}^{N}\left( {\bf \dot{r}}_{j}^{2}-2\omega _{c}x_{j}%
\dot{y}_{j}\right) -V_{1}-V_{2},
\end{equation}
where $\omega _{c}$ is the cyclotron frequency and 
\begin{equation}
V_{1}=\frac{\Omega ^{2}}{2}\sum_{j=1}^{N}{\bf r}_{j}^{2}\text{ \quad and
\quad }V_{2}=\pm \frac{\omega ^{2}}{4}\sum_{j,l=1}^{N}\left( {\bf r}_{j}-%
{\bf r}_{l}\right) ^{2}.
\end{equation}
It is obvious that the two-body potential is either attractive or repulsive
depending on the plus sign or the minus sign considered in $V_{2}$. The
magnetic field introduces a coupling in the plain perpendicular to its
direction. This means that we can separate the Lagrangian into two
contributions $L=L_{xy}+L_{z}.$ The Lagrangian $L_{z}$ simply describes a
harmonic oscillator, whereas $L_{xy}$ contains the magnetic field. The
Lagrangian can be rewritten in terms of the center of mass coordinate ${\bf R%
}\left( X,Y,Z\right) $ and the coordinates ${\bf \eta }_{j}\left(
u_{j},v_{j},w_{j}\right) $ describing the coordinates of the particles
measured from the center of mass 
\begin{equation}
{\bf R=}\frac{1}{N}\sum_{j=1}^{N}{\bf r}_{j},\text{ \qquad }{\bf \eta }_{j}=%
{\bf r}_{j}-{\bf R,}
\end{equation}
from which 
\begin{equation}
V_{1}+V_{2}=V_{CM}+V,\text{\quad }V_{CM}=\frac{1}{2}N\Omega ^{2}{\bf R}^{2},%
\text{ \quad }V=\frac{w^{2}}{2}\sum_{j=1}^{N}{\bf \eta }_{j}^{2},
\end{equation}
with 
\begin{equation}
w=\sqrt{\Omega ^{2}\pm N\omega ^{2}}.
\end{equation}
For a repulsive two-particle potential the internal frequency $w$ has to
satisfy the stability condition that the confining potential has to be
sufficiently strong to overcome the repulsion between the particles. We draw
attention to the fact that the transformation to the center of mass
coordinate system diagonalizes neither the Lagrangian nor the Hamiltonian,
because of the subsidiary condition 
\begin{equation}
\sum_{j=1}^{N}\left( {\bf r}_{j}-{\bf R}\right) =0.
\label{boundary condition}
\end{equation}

We obtain the propagator for distinguishable (indicated by a subscript $D$)
particles from the action expressed in the imaginary time variable $\beta
=1/kT$ and it is written as 
\begin{equation}
K_{D}\left( {\bf r}_{1}^{\prime \prime },\ldots ,{\bf r}_{N}^{\prime \prime
},\beta \mid {\bf r}_{1}^{\prime },\ldots ,{\bf r}_{N}^{\prime },0\right)
=K_{D}\left( \left( \bar{x}^{\prime \prime },\bar{y}^{\prime \prime }\right)
,\beta \mid \left( \bar{x}^{\prime },\bar{y}^{\prime }\right) ,0\right)
\times K_{D}\left( \bar{z}^{\prime \prime },\beta \mid \bar{z}^{\prime
},0\right) ,
\end{equation}
where the vector $\bar{x}$ denotes the $N$ dimensional $x$ coordinates of
the particles, with the notation $\bar{x}^{T}=\left( x_{1},x_{2},\ldots
,x_{N}\right) ,$ and similarly for $\bar{y}$ and $\bar{z}$. The propagator
for a single oscillator with frequency $\varpi $ in a magnetic field is well
known \cite{PRE1} and given by 
\begin{eqnarray}
K_{\omega _{L}}^{\left( 1\right) }\left( {\bf r}^{\prime \prime },\beta \mid 
{\bf r}^{\prime },0\right) &=&\sqrt{\frac{\varpi }{2\pi \sinh \beta \varpi }}%
\frac{s}{2\pi \sinh \beta s}  \nonumber  \label{propagator HO in B-veld} \\
&&\times \exp \left\{ -\frac{\varpi }{2\sinh \beta \varpi }\left[ \left(
\left( z^{\prime }\right) ^{2}+\left( z^{\prime \prime }\right) ^{2}\right)
\cosh \beta \varpi -2z^{\prime }z^{\prime \prime }\right] \right\}  \nonumber
\\
&&\times \exp \left\{ -\frac{s}{2}\frac{\left( \left( x^{\prime \prime
}\right) ^{2}+\left( y^{\prime \prime }\right) ^{2}+\left( x^{\prime
}\right) ^{2}+\left( y^{\prime }\right) ^{2}\right) \cosh \beta s-2\left(
x^{\prime }x^{\prime \prime }+y^{\prime }y^{\prime \prime }\right) \cosh
\beta \omega _{L}}{\sinh \beta s}\right\}  \nonumber \\
&&\times \exp \left\{ -i\left( \omega _{L}\left( x^{\prime \prime }y^{\prime
\prime }-x^{\prime }y^{\prime }\right) -s\frac{\sinh \beta \omega _{L}}{%
\sinh \beta s}\left( y^{\prime }x^{\prime \prime }-y^{\prime \prime
}x^{\prime }\right) \right) \right\} ,
\end{eqnarray}
where $\omega _{L}=\frac{\omega _{c}}{2}$ is the Larmor frequency and the
eigenfrequency $s$ is given by 
\begin{equation}
s=\sqrt{\varpi ^{2}+\omega _{L}^{2}}.
\end{equation}
The propagator for $N$ {\sl distinguishable} interacting oscillators in a
magnetic field thus becomes 
\begin{eqnarray}
K_{D}\left( {\bf \bar{r}}^{\prime \prime },\beta \mid {\bf \bar{r}}^{\prime
},0\right) &=&\frac{K_{\Omega }\left( \sqrt{N}Z^{\prime \prime },\beta \mid 
\sqrt{N}Z^{\prime },0\right) }{K_{w}\left( \sqrt{N}Z^{\prime \prime },\beta
\mid \sqrt{N}Z^{\prime },0\right) }\frac{K_{\omega _{L},s_{CM}}\left( \sqrt{N%
}X^{\prime \prime },\sqrt{N}Y^{\prime \prime },\beta \mid \sqrt{N}X^{\prime
},\sqrt{N}Y^{\prime },0\right) }{K_{\omega _{L},s}\left( \sqrt{N}X^{\prime
\prime },\sqrt{N}Y^{\prime \prime },\beta \mid \sqrt{N}X^{\prime },\sqrt{N}%
Y^{\prime },0\right) }  \nonumber \\
&&\times \prod_{j=1}^{N}K_{\omega _{L},s}\left( x_{j}^{\prime \prime
},y_{j}^{\prime \prime },\beta \mid x_{j}^{\prime },y_{j}^{\prime },0\right)
K_{w}\left( z_{j}^{\prime \prime },\beta \mid z_{j}^{\prime },0\right) ,
\label{propagator}
\end{eqnarray}
where 
\begin{equation}
s=\sqrt{w^{2}+\omega _{L}^{2}}\text{ \quad and \quad }s_{CM}=\sqrt{\Omega
^{2}+\omega _{L}^{2}}.
\end{equation}
The factor $\sqrt{N}$ in the center-of-mass coordinates in (\ref{propagator}%
) describes the mass $N$ (in atomic units) of the center. The denominator in
(\ref{propagator}) accounts for the fact that the internal degrees of
freedom are linearly dependent because of the subsidiary conditions (\ref
{boundary condition}). Intuitively this factor is quite natural, because the
propagator would be the product of one-particle propagators \cite{PRE1} if
the particles were independent.

Knowing the propagator for distinguishable particles, the symmetrized
density matrix $K_{I}$ for {\sl identical} particles can be obtained through
the appropriate symmetric or antisymmetric projection 
\begin{equation}
K_{I}\left( {\bf \bar{r}}^{\prime \prime },\beta \mid {\bf \bar{r}}^{\prime
},0\right) =\frac{1}{N!}\sum_{P}\xi ^{p}K_{D}\left( P{\bf \bar{r}}^{\prime
\prime },\beta \mid {\bf \bar{r}}^{\prime },0\right) ,
\end{equation}
where $P$ denotes the permutation matrix, with $\xi =+1$ for bosons and $\xi
=-1$ for fermions. Even for this harmonic model, with or without magnetic
field, the sum over the permutations has to remain rather formal at the
level of the propagator. However, for the partition function 
\begin{equation}
Z_{I}\left( \beta ,N\right) =\int d{\bf \bar{r}}K_{I}\left( {\bf \bar{r}}%
,\beta \mid {\bf \bar{r}},0\right) =\frac{1}{N!}\sum_{P}\xi ^{p}\int d{\bf 
\bar{r}}K_{D}\left( P{\bf \bar{r}},\beta \mid {\bf \bar{r}},0\right) ,
\end{equation}
analytical progress can be made with this summation as will be discussed in
the next section. First of all, one has to deal with the center-of-mass
contribution to the propagator.\ Afterwards, the summation over all possible
permutations will be rewritten as a summation over all possible cycles.

\section{Generating function of the partition function}

The center of mass is not independent of the positions of the other
particles, which complicates the calculation of the trace of the propagator.
To deal with the contribution of the center-of-mass coordinate ${\bf R}$ to
the propagator, we introduce the delta function $\delta \left( {\bf R}-\frac{%
1}{N}\sum_{j=1}^{N}{\bf r}_{j}\right) $ in its Fourier representation as in 
\cite{PRE1}. This delta function allows one to formally treat the
center-of-mass coordinate as an independent variable. Applying this identity
to the partition function, one ends up with 
\begin{eqnarray}
Z_{I}\left( \beta ,N\right) &=&\int d{\bf R}\int \frac{d{\bf k}}{\left( 2\pi
\right) ^{3}}e^{i{\bf k\cdot R}}\frac{K_{\Omega }\left( \sqrt{N}Z,\beta \mid 
\sqrt{N}Z,0\right) }{K_{w}\left( \sqrt{N}Z,\beta \mid \sqrt{N}Z,0\right) }%
\frac{K_{\omega _{L,}s_{CM}}\left( \sqrt{N}X,\sqrt{N}Y,\beta \mid \sqrt{N}X,%
\sqrt{N}Y,0\right) }{K_{\omega _{L,}s}\left( \sqrt{N}X,\sqrt{N}Y,\beta \mid 
\sqrt{N}X,\sqrt{N}Y,0\right) }  \nonumber
\label{Partitiefunctie met identiteit} \\
&&\times \int d{\bf \bar{r}}\frac{1}{N!}\sum_{P}\xi
^{p}\prod_{j=1}^{N}K_{\omega _{L,}s}\left( \left( Px\right) _{j},\left(
Py\right) _{j},\beta \mid x_{j},y_{j},0\right) K_{w}\left( \left( Pz\right)
_{j},\beta \mid z_{j},0\right) e^{-i\vec{k}\cdot \vec{r}_{j}/N}.
\end{eqnarray}
The problem at hand is the rewriting of the summation over the permutations
as a sum over all possible cycles \cite{PRE1}. This cyclic decomposition
requires the solution of the path integral for a driven harmonic oscillator
in a magnetic field, which is discussed in appendix A.

A permutation can be decomposed into $M_{\ell }$ cycles of length $\ell $,
and the positive integers $M_{\ell }$ and $\ell $ have to satisfy the
constraint 
\begin{equation}
\sum_{\ell }\ell M_{\ell }=N.  \label{cyclicondition}
\end{equation}
The number $M\left( M_{1},\ldots ,M_{N}\right) $ of permutations with $M_{1}$
cycles of length $1$,$\ldots $, $M_{\ell }$ cycles of length $\ell $,$\ldots 
$ is given by $M\left( M_{1},\ldots ,M_{N}\right) =N!/\left[ \prod_{\ell
}M_{\ell }!\ell ^{M_{\ell }}\right] .$ Furthermore, a cycle of length $\ell $
will be obtained from $\ell -1$ permutations. Thus, the sign factor $\xi
^{p} $ can be rewritten as $\xi ^{p}=\prod_{\ell }\xi ^{\left( \ell
-1\right) M_{\ell }}.$ These considerations enable one to rewrite the
partition function as 
\begin{eqnarray}
Z_{I}\left( \beta ,N\right) &=&\int d{\bf R}\int \frac{d{\bf k}}{\left( 2\pi
\right) ^{3}}e^{i{\bf k\cdot R}}\frac{K_{\Omega }\left( \sqrt{N}Z,\beta \mid 
\sqrt{N}Z,0\right) }{K_{w}\left( \sqrt{N}Z,\beta \mid \sqrt{N}Z,0\right) }%
\frac{K_{\omega _{L,}s_{CM}}\left( \sqrt{N}X,\sqrt{N}Y,\beta \mid \sqrt{N}X,%
\sqrt{N}Y,0\right) }{K_{\omega _{L,}s}\left( \sqrt{N}X,\sqrt{N}Y,\beta \mid 
\sqrt{N}X,\sqrt{N}Y,0\right) }  \nonumber \\
&&\times \sum_{M_{1},\ldots ,M_{N}}\prod_{j=1}^{N}\frac{\xi ^{\left( \ell
-1\right) M_{\ell }}}{M_{\ell }!\ell ^{M_{\ell }}}\left[ {\cal K}_{\ell
}\left( {\bf k}\right) \right] ^{M_{\ell }},
\end{eqnarray}
with 
\begin{equation}
{\cal K}_{\ell }\left( {\bf k}\right) =\int d{\bf r}_{\ell +1}\ldots \int d%
{\bf r}_{1}\delta \left( {\bf r}_{\ell +1}-{\bf r}_{1}\right)
\prod_{j=1}^{\ell }K_{\omega _{L,}s}\left( x_{j+1},y_{j+1},\beta \mid
x_{j},y_{j},0\right) K_{w}\left( z_{j+1},\beta \mid z_{j},0\right) e^{-i\vec{%
k}\cdot \vec{r}_{j}/N}.
\end{equation}
The delta function explicitly indicates that the trace is taken over a cycle
of length $\ell $. It is obvious that ${\cal K}_{\ell }\left( {\bf k}\right) 
$ factorizes as 
\begin{equation}
{\cal K}_{\ell }\left( {\bf k}\right) ={\cal K}_{\ell }\left(
k_{x},k_{y}\right) {\cal K}_{\ell }\left( k_{z}\right) .
\end{equation}
Taking into account the semigroup property of the propagators $K_{\omega
_{L},s}\left( x_{j+1},y_{j+1},\beta \mid x_{j},y_{j},0\right) $ and $%
K_{w}\left( z_{j+1},\beta \mid z_{j},0\right) $, one immediately recognizes
in ${\cal K}_{\ell }\left( {\bf k}\right) $ the partition function of a
driven harmonic oscillator in a magnetic field, with the driving force 
\begin{equation}
{\bf f}\left( \tau \right) =i\frac{{\bf k}}{N}\sum_{j=0}^{\ell -1}\delta
\left( \tau -j\beta \right) .
\end{equation}
The calculation of the propagator and the partition function for a driven
harmonic oscillator in a magnetic field, with the Lagrangian 
\begin{equation}
L_{{\bf f},\omega _{L}}^{\left( 1\right) }=\frac{1}{2}\left( \dot{x}^{2}+%
\dot{y}^{2}\right) -2\omega _{L}x\dot{y}-\frac{w^{2}}{2}\left(
x^{2}+y^{2}\right) +f_{x}\left( \tau \right) x+f_{y}\left( \tau \right) y,
\end{equation}
is illustrated in appendix A. As mentioned above, the hard core of this
approach is the evaluation of the classical action, but a fully stochastic
method \cite{Simon} could as well have been followed. After tedious algebra
one eventually finds for the partition function 
\begin{eqnarray}
Z_{I}\left( \beta ,N\right) &=&\int d{\bf R}\frac{K_{\Omega }\left( \sqrt{N}%
Z,\beta \mid \sqrt{N}Z,0\right) }{K_{w}\left( \sqrt{N}Z,\beta \mid \sqrt{N}%
Z,0\right) }\frac{K_{\omega _{L,}s_{CM}}\left( \sqrt{N}X,\sqrt{N}Y,\beta
\mid \sqrt{N}X,\sqrt{N}Y,0\right) }{K_{\omega _{L,}s}\left( \sqrt{N}X,\sqrt{N%
}Y,\beta \mid \sqrt{N}X,\sqrt{N}Y,0\right) }  \nonumber \\
&&\times \int \frac{d{\bf k}}{\left( 2\pi \right) ^{3}}e^{i{\bf k\cdot R}%
}\exp \left( -\frac{1}{4s}\frac{k_{x}^{2}+k_{y}^{2}}{N}\frac{\sinh \beta s}{
\cosh \beta s-\cosh \beta \omega _{L}}-\frac{1}{4w}\frac{k_{z}^{2}}{N}\frac{%
\sinh \beta w}{\cosh \beta w-1}\right)  \nonumber \\
&&\times \sum_{M_{1},\ldots ,M_{N}}\prod_{j=1}^{N}\frac{\xi ^{\left( \ell
-1\right) M_{\ell }}}{M_{\ell }!\ell ^{M_{\ell }}}\left( \frac{1}{8\sinh 
\frac{\ell \beta \left( s+\omega _{L}\right) }{2}\sinh \frac{\ell \beta
\left( s-\omega _{L}\right) }{2}\sinh \frac{\ell \beta w}{2}}\right)
^{M_{\ell }}.
\end{eqnarray}
The remaining integrations over ${\bf k}$ and ${\bf R}$ are Gaussian and
relatively easy to perform, leading to 
\begin{equation}
Z_{I}\left( \beta ,N\right) =\frac{\sinh \frac{\beta \left( s+\omega
_{l}\right) }{2}\sinh \frac{\beta \left( s-\omega _{l}\right) }{2}\sinh 
\frac{\beta w}{2}}{\sinh \frac{\beta \left( s_{CM}+\omega _{l}\right) }{2}%
\sinh \frac{\beta \left( s_{CM}-\omega _{l}\right) }{2}\sinh \frac{\beta
\Omega }{2}}{\Bbb Z}_{I}\left( N\right) ,
\label{part. func. in cyclic decomp.}
\end{equation}
with 
\begin{equation}
{\Bbb Z}_{I}\left( \beta ,N\right) =\sum_{M_{1},\ldots ,M_{N}}\prod_{j=1}^{N}%
\frac{\xi ^{\left( \ell -1\right) M_{\ell }}}{M_{\ell }!\ell ^{M_{\ell }}}%
\left( \frac{1}{8\sinh \frac{\ell \beta \left( s+\omega _{L}\right) }{2}%
\sinh \frac{\ell \beta \left( s-\omega _{L}\right) }{2}\sinh \frac{\ell
\beta w}{2}}\right) ^{M_{\ell }}.
\end{equation}

The contribution ${\Bbb Z}_{I}\left( \beta ,N\right) $ derives from the
internal degrees of freedom, treated as independent particles. It contains
the full influence of the statistics of the particles, and leads to the true
partition function $Z_{I}\left( \beta ,N\right) $ by multiplication with a
simple analytical factor. In practice, the condition (\ref{cyclicondition})
complicates the use of the above expression for the partition function for a
large number of particles. However this difficulty can be overcome through
the use of the generating function. From the generating function one can
then extract the partition function through an inversion of its defining
Taylor series.

\subsection{Generating function and recurrence relation for the partition
function}

The generating function technique was used before \cite{PRE1} to obtain the
partition function of a set of harmonically interacting identical
oscillators in the absence of a magnetic field. In the presence of a
magnetic field, a similar construction can be used. Introducing the
generating function as 
\begin{equation}
\Xi _{I}\left( \beta ,u\right) =\sum_{N=0}^{\infty }{\Bbb Z}_{I}\left(
N\right) u^{N},  \label{Taylor-series for the gen. func.}
\end{equation}
(with ${\Bbb Z}_{I}\left( \beta ,0\right) =1$ by definition), the partition
function for the internal degrees of freedom can be obtained from 
\begin{equation}
{\Bbb Z}_{I}\left( \beta ,N\right) =\frac{1}{N!}\frac{d^{N}}{du^{N}}\left.
\Xi _{I}\left( u\right) \right| _{u=0}.  \label{partition - chain rule}
\end{equation}
The generating function itself can be obtained with straightforward algebra 
\begin{equation}
\Xi _{I}\left( \beta ,u\right) =\exp \left( \sum_{\ell =1}^{\infty }\frac{%
\xi ^{\ell -1}}{\ell }\frac{\left( b_{1}b_{2}b\right) ^{\frac{1}{2}\ell
}u^{\ell }}{\left( 1-b^{\ell }\right) \left( 1-b_{1}^{\ell }\right) \left(
1-b_{2}^{\ell }\right) }\right) ,
\end{equation}
with the following notation 
\begin{equation}
b=e^{-\beta w},\text{ \qquad }b_{1}=e^{-\beta \left( s+\omega _{L}\right) },%
\text{ \qquad }b_{2}=e^{-\beta \left( s-\omega _{L}\right) }.
\end{equation}
The cyclic summation can be rewritten in terms of the occupation number
representation which directly involves the single particle energy levels: 
\begin{equation}
\Xi _{I}\left( \beta ,u\right) =\prod_{\nu ,\nu _{1},\nu _{2}=0}^{\infty
}\left( 1-\xi ub_{1}^{\nu _{1}+\frac{1}{2}}b_{2}^{\nu _{2}+\frac{1}{2}%
}b^{\nu +\frac{1}{2}}\right) ^{-\xi }.  \label{gen. func.}
\end{equation}
By simply applying the chain rule, the expression (\ref{partition - chain
rule}) for the partition function can be written as a recurrence relation 
\begin{equation}
{\Bbb Z}_{I}\left( \beta ,N\right) =\frac{1}{N}\sum_{\ell =1}^{N}\xi ^{\ell
-1}\frac{\left( b_{1}b_{2}b\right) ^{\frac{\ell }{2}}}{\left( 1-b_{1}^{\ell
}\right) \left( 1-b_{2}^{\ell }\right) \left( 1-b^{\ell }\right) }{\Bbb Z}
_{I}\left( \beta ,N-\ell \right) .
\end{equation}
However if the number of particles increases, this recurrence relation
becomes numerically unpractical because of a numerical sign problem for
fermions and drastically increasing simulation time for bosons. For the
remaining part of this paper, the attention will be focussed on the fermion
case.

\section{Thermodynamical properties}

The thermodynamical properties of the fermion model can in essence be
determined from the contribution ${\Bbb Z}_{F}\left( \beta ,N\right) $ of
the internal degrees of freedom. As is clear from (\ref{part. func. in
cyclic decomp.}), the center of mass correction only adds a trivial
contribution to the free energy. All the effects of the fermion statistics
are collected in ${\Bbb Z}_{F}\left( \beta ,N\right) .$ We first study the
zero temperature limit, in which special attention will be paid to the two
dimensional case in the $xy$-plane, and subsequently the evolution of the
free energy and the internal energy as a function of the temperature and of
the magnetic field.

\subsection{Zero temperature limit}

The ground state properties of the fermion model crucially depend on the
single-particle energy levels $E_{\nu ,\nu _{1},\nu _{2}}$. These levels and
their occupation by fermions have been discussed and plotted earlier, e.g.,
in Ref. \cite{Kouwenhove}. For easier reference, we plot the 20 lowest
levels Fig. 1. To guide the eye, the Fermi energies $E_{F}$ corresponding to
the fully occupied levels at $\omega _{L}=0$ are indicated by the dashed
line. Note that the magnetic field does not substantially influence the
magnitude of the Fermi energy, which remains of order $N^{1/3}$ for
sufficiently large $N$. The magnetic field immediately lifts the degeneracy,
but with increasing magnetic field other degeneracies appear and disappear
again at particular values of the magnetic field. Although these
degeneracies have little effect on the magnitude of the ground state energy $%
E_{G}=\sum_{E<E_{F}}E$, they have a drastic effect on the magnetic
susceptibility, which is proportional to $dE_{G}/d\omega _{L}$, as shown in
Fig. 2 as a function of the magnetic field. The discontinuities in the
magnetic susceptibility occur at those values of the magnetic field where
the single-particle energies become degenerate.

\subsection{Free energy and internal energy}

As mentioned above, the sign problem for fermions can be worked around by
inverting the defining Taylor-series (\ref{Taylor-series for the gen. func.}%
) for the generating function. The Fowler-Darwin method \cite{Kubo} provides
an accurate and elegant way \cite{Kringintegraal} to realize this inversion: 
\begin{equation}
{\Bbb Z}_{F}\left( \beta ,N\right) =\frac{1}{2\pi i}\oint_{C}\frac{\Xi
_{F}\left( \beta ,z\right) }{z^{N+1}}dz.  \label{Contour integral}
\end{equation}
If one considers a circular contour $z=ue^{i\theta }$ with radius $u$, an
optimal value of $u$ can be determined by the method of steepest descent 
\begin{equation}
\frac{d}{du}\left( \ln \Xi _{F}\left( \beta ,u\right) -N\ln u\right)
=0\Longrightarrow N=u\frac{d}{du}\ln \Xi _{F}\left( \beta ,u\right) .
\end{equation}
Using (\ref{gen. func.}), this condition becomes 
\begin{equation}
N=\sum_{\nu ,\nu _{1},\nu _{2}=0}^{\infty }n_{\nu ,\nu _{1},\nu _{2}},
\label{N}
\end{equation}
with 
\begin{equation}
n_{\nu ,\nu _{1},\nu _{2}}=\frac{1}{1+e^{\beta E_{\nu ,\nu _{1},\nu _{2}}}u},%
\text{ \qquad }E_{\nu ,\nu _{1},\nu _{2}}=\left( \nu _{1}+\frac{1}{2}\right)
s_{1}+\left( \nu _{2}+\frac{1}{2}\right) s_{2}+\left( \nu +\frac{1}{2}%
\right) w.
\end{equation}
If $u$ would be interpreted as the fugacity $u=e^{\beta \mu }$ with chemical
potential $\mu $, one thus would recover similar results as for the
expectation value of the number of particles in the grand canonical
ensemble. The result for the chemical potential as a function of temperature
is shown in Fig. 3. for various values of the magnetic field and for $N=2$.
The chemical potential is plotted in units of the chemical potential at $T=0$%
, and the temperature in units of $wN^{1/3}$, which is the order of
magnitude for the Fermi energy. In Fig. 4 and 5 the corresponding results
are shown for $N=10$ and $N=100$. For $N\gtrsim 100$ it turns out that $\mu
\left( T\right) /\mu \left( T=0\right) $ as a function of $kT/wN^{1/3}$
becomes almost independent of both the number of particles and of the
magnetic field.

However, in the present treatment the determination of $u=e^{\beta \mu }$
from (\ref{N}) only gives the zero-order contribution to the partition
function. A correction by the integration factor in (\ref{Contour integral})
has to be applied. Using the symmetry of the integrand in (\ref{Contour
integral}), the partition function can be rewritten as 
\begin{equation}
{\Bbb Z}_{F}\left( \beta ,N\right) =\frac{\Xi _{F}\left( \beta ,u\right) }{%
u^{N}}\frac{1}{2\pi }\int_{0}^{2\pi }\frac{\Xi _{F}\left( \beta ,ue^{i\theta
}\right) }{\Xi _{F}\left( \beta ,u\right) }e^{-iN\theta }d\theta =\frac{\Xi
_{F}\left( \beta ,u\right) }{u^{N}}\int_{0}^{\pi }\Psi \left( \theta \right)
d\theta ,
\end{equation}
with 
\begin{equation}
\Psi \left( \theta \right) =%
\mathop{\rm Re}%
\left[ \frac{1}{\pi }e^{-i\theta N}\frac{\Xi _{F}\left( \beta ,ue^{i\theta
}\right) }{\Xi _{F}\left( \beta ,u\right) }\right] .
\end{equation}
The function $\Psi \left( \theta \right) $ has to be calculated and
integrated numerically. The determination of the free energy 
\begin{equation}
{\Bbb F}_{F}\left( \beta ,N\right) =-\frac{1}{N}\ln {\Bbb Z}_{F}\left( \beta
,N\right) ={\Bbb F}_{F}^{\left( 0\right) }\left( \beta ,N\right) -\frac{1}{%
\beta }\ln \int_{0}^{\pi }\Psi \left( \theta \right) d\theta
\end{equation}
then becomes straightforward, with the zero-order contribution ${\Bbb F}%
_{F}^{\left( 0\right) }\left( \beta ,N\right) =-\frac{1}{\beta }\ln \frac{
\Xi _{F}\left( \beta ,u\right) }{u^{N}}$ from the steepest-descent
approximation. In the absence of a magnetic field, the results are discussed
in \cite{PRE4}. For $\omega _{L}=2w$ the free energy per particle in units
of the fermi energy is plotted as a function of $kT/E_{F}$ for 10 fermions
in Fig. 6 and for 100 fermions in Fig. 7, and compared to the zero-order
steepest descent contribution. Again, for $N\gtrsim 100$ the finite number
corrections upon the thermodynamical limit become negligible for all
practical purposes. The internal energy ${\Bbb U}_{F}=\frac{d}{d\beta }%
\left( \beta {\Bbb F}_{F}\right) $ shows the same universality, as is shown
in Fig. 8 where the internal energy per particle in units of the Fermi
energy is plotted versus the temperature in units of the Fermi temperature
for $\omega _{L}=0,w$ and $2w$.

\section{Conclusion and discussion}

Using the path integral approach of symmetrized density matrices for
identical particles, the thermodynamical properties were calculated for a
three dimensional model of $N$ harmonically interacting spin-polarized
fermions in a parabolic potential well in the presence of a magnetic field.
The method used is a generalization of the procedure developed earlier in
the absence of a magnetic field. Explicit results were obtained for the
ground state energy, the free energy and the internal energy for a limited
number of particles. The model can be described as a number of spin
polarized identical particles in a parabolic confinement potential
interacting through a special many body interaction with the consequence
that the center of mass is allowed to move independent from the other
degrees of freedom. For an analoguous model other forms of confinement
potentials have been investigated without two body interaction. \cite
{VVIGPRE98}

The statistics with a finite number of particles in the confinement
potential and the cross over to density dependent expressions known from the
thermodynamical limit can be studied in this model: as soon as the number of
fermions is sufficiently large (in the order of $N\gtrsim 100$) the results
are shown to agree and the finite number corrections become relatively
small. The internal energy turns out to obey a scaling law, similar to the
scaling from the continuum approximation for the density of states.

\acknowledgements %
%
Part of this work is performed in the framework of the FWO projects No.
1.5.729.94, 1.5.545.98, G.0287.95, G.0071.98, and WO.073.94N
(Wetenschappelijke Onderzoeksgemeenschap over ``Laag-dimensionele
systemen''), the ``Interuniversitaire Attractiepolen - Belgische Staat,
Diensten van de Eerste Minister - Wetenschappelijke, Technische en Culturele
aangelegenheden'', and in the framework of the BOF\ NOI 1997 projects of the
Universiteit Antwerpen. One of the authors (F.B.) acknowledges the FWO
(Fonds voor Wetenschappelijk Onderzoek - Vlaanderen) for financial support.
S.F. acknowledges the University of Antwerpen (UIA) for a research grant.

\appendix

\section{The path integral of the model for distinguishable particles in the
presence of a magnetic field, and a time-dependent driving force.}

The propagator of a two dimensional harmonic oscillator in the presence of a
magnetic field, characterized by the Larmor frequency $\omega _{L}=\omega
_{c}/2$, and under the influence of a time-dependent driving force ${\bf f}%
=\left( f_{x},f_{y}\right) $, provides the basic building blocks for the
harmonic model system of identical interacting particles which is the
subject of the present paper. Although the calculation of this propagator
relies on standard techniques, to the best of our knowledge it is not
documented in the literature. Therefore we discuss its derivation here in
some detail. The Lagrangian under consideration is given by (in atomic units 
$\hbar =m=\left| e\right| =1$) 
\begin{equation}
L_{{\bf f},\omega _{L}}^{\left( 1\right) }=\frac{1}{2}\left( \dot{x}^{2}+%
\dot{y}^{2}\right) -2\omega _{L}x\dot{y}-\frac{w^{2}}{2}\left(
x^{2}+y^{2}\right) +f_{x}\left( \tau \right) x+f_{y}\left( \tau \right) y.
\end{equation}
The {\sl classical }equations of motion in Euclidean time $\tau =it$ are 
\begin{eqnarray}
-\frac{d^{2}x}{d\tau ^{2}} &=&-2i\omega _{L}\frac{dy}{d\tau }-w^{2}x+f_{x},
\\
-\frac{d^{2}y}{d\tau ^{2}} &=&2i\omega _{L}\frac{dx}{d\tau }-w^{2}y+f_{y}.
\end{eqnarray}
This set of coupled differential equations can be solved and yields 
\begin{equation}
x\left( \tau \right) =x_{h}\left( \tau \right) +x_{p}\left( \tau \right) ,%
\text{ \qquad }y\left( \tau \right) =y_{h}\left( \tau \right) +y_{p}\left(
\tau \right) .
\end{equation}
The solutions $x_{h}\left( \tau \right) $ and $y_{h}\left( \tau \right) $ of
the homogeneous equations of motion (without the driving force), which
exhaust the boundary conditions $x_{h}\left( 0\right) =x^{\prime },$ $%
x_{h}\left( \beta \right) =x^{\prime \prime },$ $y_{h}\left( 0\right)
=y^{\prime },$ $y_{h}\left( \beta \right) =y^{\prime \prime }$, are found to
be 
\begin{eqnarray}
\left( 
\begin{array}{c}
x_{h}\left( \tau \right) \\ 
y_{h}\left( \tau \right)
\end{array}
\right) &=&\frac{\sinh s\left( \beta -\tau \right) }{\sinh \beta s}\left( 
\begin{array}{cc}
\cosh \omega _{L}\tau & i\sinh \omega _{L}\tau \\ 
-i\sinh \omega _{L}\tau & \cosh \omega _{L}\tau
\end{array}
\right) \left( 
\begin{array}{c}
x^{\prime } \\ 
y^{\prime }
\end{array}
\right)  \nonumber \\
&&+\frac{\sinh s\tau }{\sinh \beta s}\left( 
\begin{array}{cc}
\cosh \omega _{L}\left( \beta -\tau \right) & -i\sinh \omega _{L}\left(
\beta -\tau \right) \\ 
i\sinh \omega _{L}\left( \beta -\tau \right) & \cosh \omega _{L}\left( \beta
-\tau \right)
\end{array}
\right) \left( 
\begin{array}{c}
x^{\prime \prime } \\ 
y^{\prime \prime }
\end{array}
\right) ,
\end{eqnarray}
with 
\begin{equation}
s=\sqrt{\omega _{L}^{2}+w^{2}}.
\end{equation}
The derivation of the particular solutions $x_{p}\left( \tau \right) $ and $%
y_{p}\left( \tau \right) ,$ with the boundary conditions $x_{p}\left(
0\right) =x_{p}\left( \beta \right) =y_{p}\left( 0\right) =y_{p}\left( \beta
\right) =0$, is slightly more involved but eventually results in 
\begin{eqnarray}
\left( 
\begin{array}{c}
x_{p}\left( \tau \right) \\ 
y_{p}\left( \tau \right)
\end{array}
\right) &=&\frac{1}{s}\frac{\sinh s\left( \beta -\tau \right) }{\sinh \beta s%
}\int_{0}^{\tau }\left( 
\begin{array}{cc}
\cosh \omega _{L}\left( \tau -\sigma \right) & i\sinh \omega _{L}\left( \tau
-\sigma \right) \\ 
-i\sinh \omega _{L}\left( \tau -\sigma \right) & \cosh \omega _{L}\left(
\tau -\sigma \right)
\end{array}
\right) \left( 
\begin{array}{c}
f_{x}\left( \sigma \right) \\ 
f_{y}\left( \sigma \right)
\end{array}
\right) \sinh s\sigma d\sigma  \nonumber \\
&&+\frac{1}{s}\frac{\sinh s\tau }{\sinh \beta s}\int_{\tau }^{\beta }\left( 
\begin{array}{cc}
\cosh \omega _{L}\left( \tau -\sigma \right) & i\sinh \omega _{L}\left( \tau
-\sigma \right) \\ 
-i\sinh \omega _{L}\left( \tau -\sigma \right) & \cosh \omega _{L}\left(
\tau -\sigma \right)
\end{array}
\right) \left( 
\begin{array}{c}
f_{x}\left( \sigma \right) \\ 
f_{y}\left( \sigma \right)
\end{array}
\right) \sinh s\left( \beta -\sigma \right) d\sigma .
\end{eqnarray}
Given the classical trajectory with initial position $\left( x^{\prime
},y^{\prime }\right) $ and final position $\left( x^{\prime \prime
},y^{\prime \prime }\right) $ after an imaginary time lapse $\beta $, the
corresponding classical action $S_{{\bf f},cl}=\int_{0}^{\beta }L_{{\bf f}%
}d\tau $ can be found by elementary methods. This eventually results in 
\begin{eqnarray}
&&S_{{\bf f},cl}\left( \left. x^{\prime \prime },y^{\prime \prime },\beta
\right| x^{\prime },y^{\prime },0\right)  \nonumber \\
&=&-\frac{s}{\sinh \beta s}\left( 
\begin{array}{c}
\frac{1}{2}\left( \left( x^{\prime \prime }\right) ^{2}+\left( y^{\prime
\prime }\right) ^{2}+\left( x^{\prime }\right) ^{2}+\left( y^{\prime
}\right) ^{2}\right) \cosh \beta s \\ 
-\left( x^{\prime }x^{\prime \prime }+y^{\prime }y^{\prime \prime }\right)
\cosh \beta \omega _{L}+i\left( x^{\prime }y^{\prime \prime }-x^{\prime
\prime }y^{\prime }\right) \sinh \beta \omega _{L}
\end{array}
\right) -i\omega _{L}\left( x^{\prime \prime }y^{\prime \prime }-x^{\prime
}y^{\prime }\right)  \nonumber \\
&&+\frac{1}{\sinh \beta s}\left( 
\begin{array}{l}
+x^{\prime }\int_{0}^{\beta }\left( f_{x}\left( \tau \right) \cosh \omega
_{L}\tau -if_{y}\left( \tau \right) \sinh \omega _{L}\tau \right) \sinh
s\left( \beta -\tau \right) d\tau \\ 
+x^{\prime \prime }\int_{0}^{\beta }\left( f_{x}\left( \tau \right) \cosh
\omega _{L}\left( \beta -\tau \right) +if_{y}\left( \tau \right) \sinh
\omega _{L}\left( \beta -\tau \right) \right) \sinh s\tau d\tau \\ 
+y^{\prime }\int_{0}^{\beta }\left( f_{y}\left( \tau \right) \cosh \omega
_{L}\tau +if_{x}\left( \tau \right) \sinh \omega _{L}\tau \right) \sinh
s\left( \beta -\tau \right) d\tau \\ 
+y^{\prime \prime }\int_{0}^{\beta }\left( f_{y}\left( \tau \right) \cosh
\omega _{L}\left( \beta -\tau \right) -if_{x}\left( \tau \right) \sinh
\omega _{L}\left( \beta -\tau \right) \right) \sinh s\tau d\tau
\end{array}
\right)  \nonumber \\
&&+\frac{1}{s\sinh \beta s}\int_{0}^{\beta }\int_{0}^{\tau }\left( 
\begin{array}{c}
\left( f_{x}\left( \tau \right) f_{x}\left( \sigma \right) +f_{y}\left( \tau
\right) f_{y}\left( \sigma \right) \right) \cosh \omega _{L}\left( \tau
-\sigma \right) \\ 
+i\left( f_{x}\left( \tau \right) f_{y}\left( \sigma \right) +f_{y}\left(
\tau \right) f_{x}\left( \sigma \right) \right) \sinh \omega _{L}\left( \tau
-\sigma \right)
\end{array}
\right) \sinh \sigma s\sinh s\left( \beta -\tau \right) d\sigma d\tau .
\end{eqnarray}

Since the Lagrangian is quadratic in the coordinates and the velocities, the
quantum mechanical propagator is determined by the classical action, apart
from a trivial normalization factor. The latter can be determined by
elementary methods. The calculation presents no difficulties and results in 
\begin{equation}
K_{{\bf f}}\left( \left. x^{\prime \prime },y^{\prime \prime },\beta \right|
x^{\prime },y^{\prime },0\right) =\frac{s}{2\pi \sinh \beta s}\exp \left[ S_{%
{\bf f},cl}\left( \left. x^{\prime \prime },y^{\prime \prime },\beta \right|
x^{\prime },y^{\prime },0\right) \right] .
\end{equation}
If one takes the limit $\omega _{L}\rightarrow 0$, the correct result \cite
{Feynman1} is recovered.

For the treatment of the cyclic summations for identical particles, it is
essential to know the trace of this propagator 
\begin{equation}
Z_{{\bf f}}=\int \int K_{{\bf f}}\left( \left. x,y,\beta \right|
x,y,0\right) dxdy.
\end{equation}
The calculation of this quantity is straightforward and after some algebra
one obtains 
\begin{equation}
Z_{{\bf f}}=\frac{1}{2\left( \cosh \beta s-\cosh \beta \omega _{L}\right) }%
\exp \left[ \frac{\Phi \left( \beta \right) }{4s\left( \cosh \beta s-\cosh
\beta \omega _{L}\right) }\right] ,
\end{equation}
with 
\begin{eqnarray}
\Phi \left( \beta \right) &=&\int_{0}^{\beta }\int_{0}^{\beta }\left(
f_{x}\left( \tau \right) f_{x}\left( \sigma \right) +f_{y}\left( \tau
\right) f_{y}\left( \sigma \right) \right) \left( 
\begin{array}{c}
\cosh \omega _{L}\left( \tau -\sigma \right) \sinh s\left( \beta -\left|
\tau -\sigma \right| \right) \\ 
+\cosh \omega _{L}\left( \beta -\left| \tau -\sigma \right| \right) \sinh
s\left( \tau -\sigma \right)
\end{array}
\right) d\sigma d\tau  \nonumber \\
&&+i\int_{0}^{\beta }\int_{0}^{\beta }\left( f_{x}\left( \tau \right)
f_{y}\left( \sigma \right) +f_{y}\left( \tau \right) f_{x}\left( \sigma
\right) \right) \left( 
\begin{array}{c}
\sinh \omega _{L}\left( \tau -\sigma \right) \sinh s\left( \beta -\left|
\sigma -\tau \right| \right) \\ 
-\sinh \omega _{L}\left( \beta -\left| \tau -\sigma \right| \right) \sinh
s\left( \tau -\sigma \right)
\end{array}
\right) d\sigma d\tau .
\end{eqnarray}
Again, if one takes the limit of a vanishing magnetic field one finds the
correct result \cite{Feynman3}.

\begin{center}
\newpage

{\bf Figure captions }
\end{center}

\begin{description}
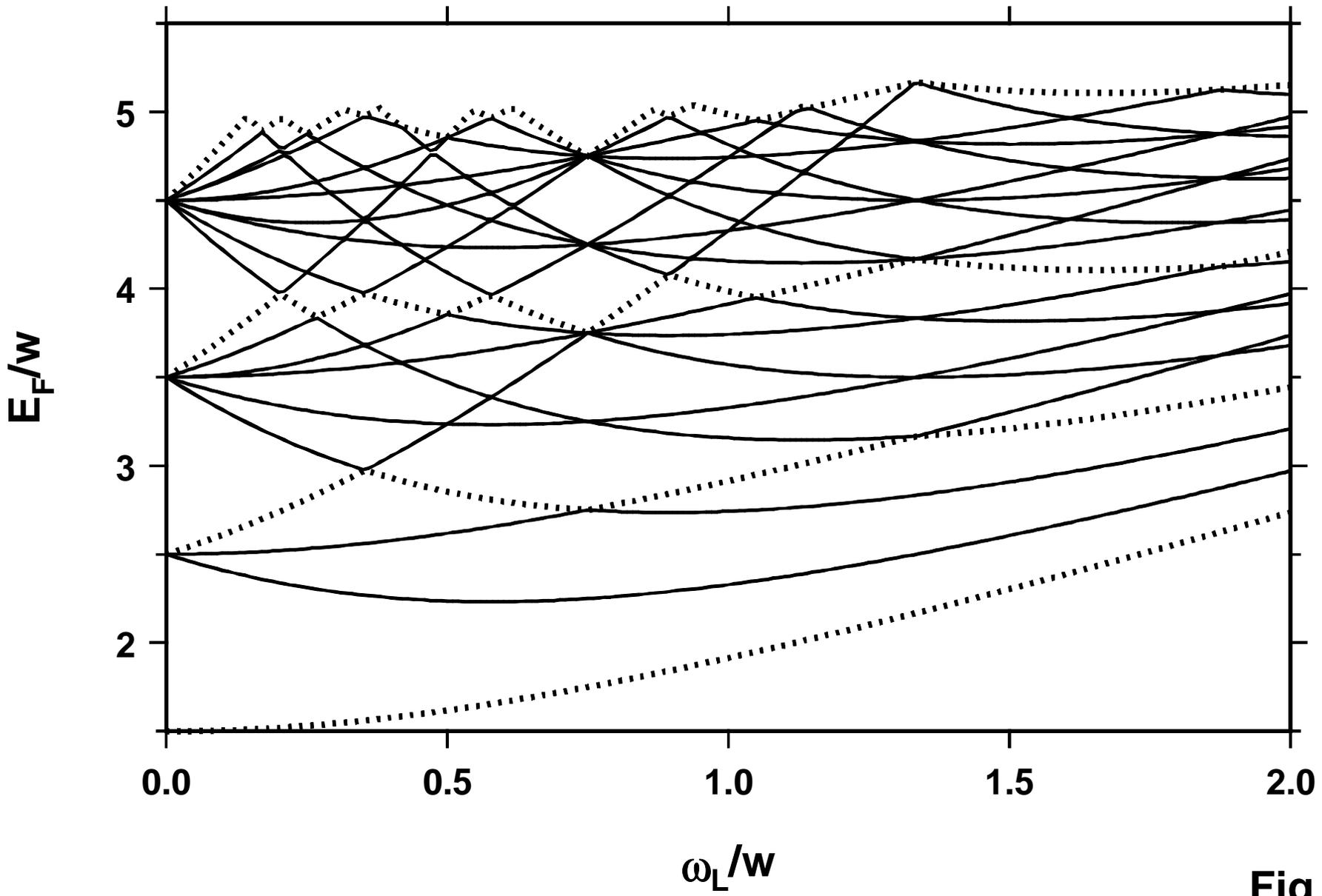
\item  {\bf Fig. 1: }Lowest single particle energy levels (in units of $w$)
as a function of the Larmor frequency. The Fermi energies corresponding to
1, 4, 10 and 20 fermions (i.e. for closed shells in the absence of a
magnetic field) are emphasized by dashed lines. The results can also be
found in Ref. \cite{Kouwenhove}.

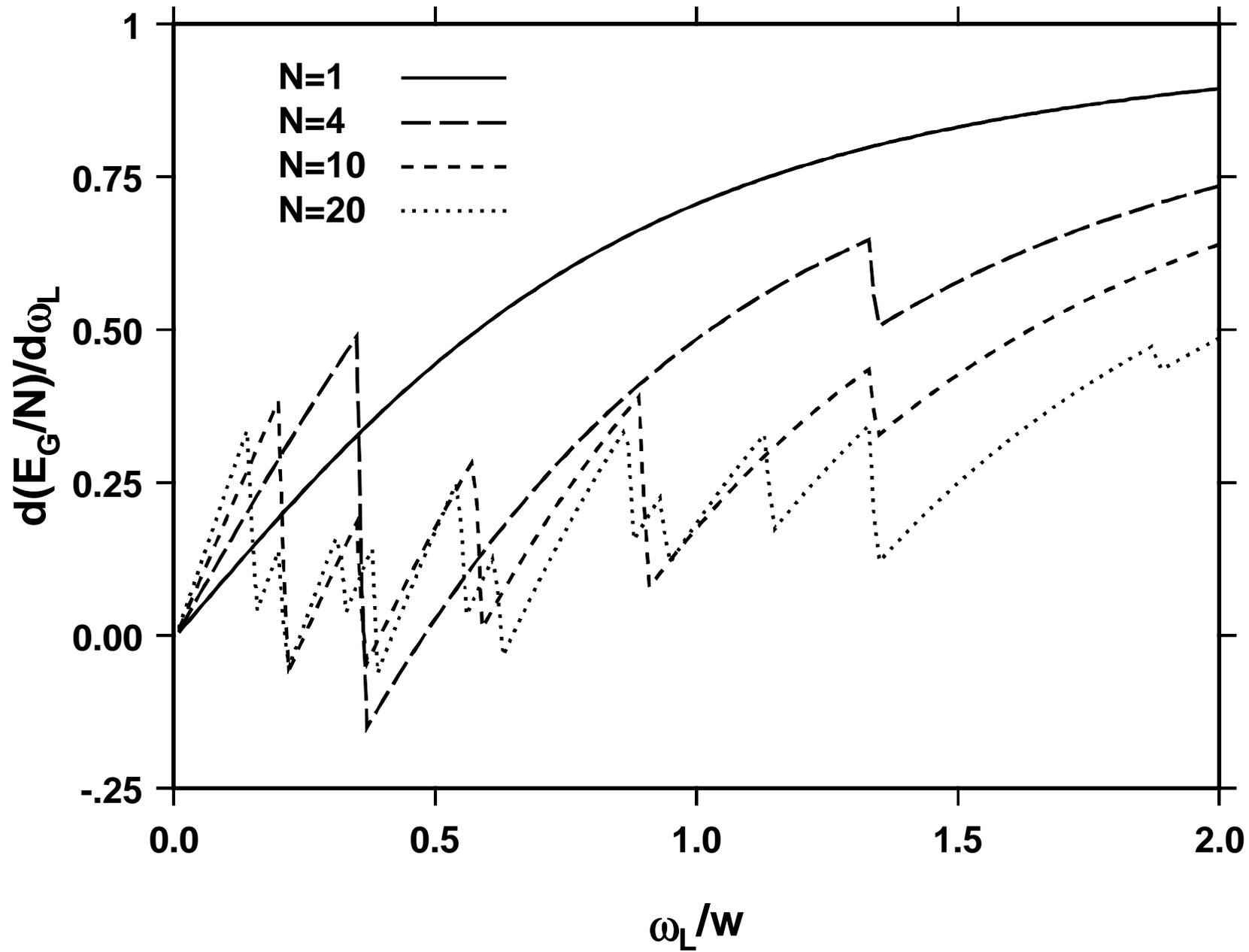
\item  {\bf Fig. 2: }Scaled magnetic susceptibility $\left( 1/N\right)
\left( dE_{G}/d\omega _{L}\right) $ as a function of the magnetic field for
1, 4, 10, 20 fermions in the ground state.

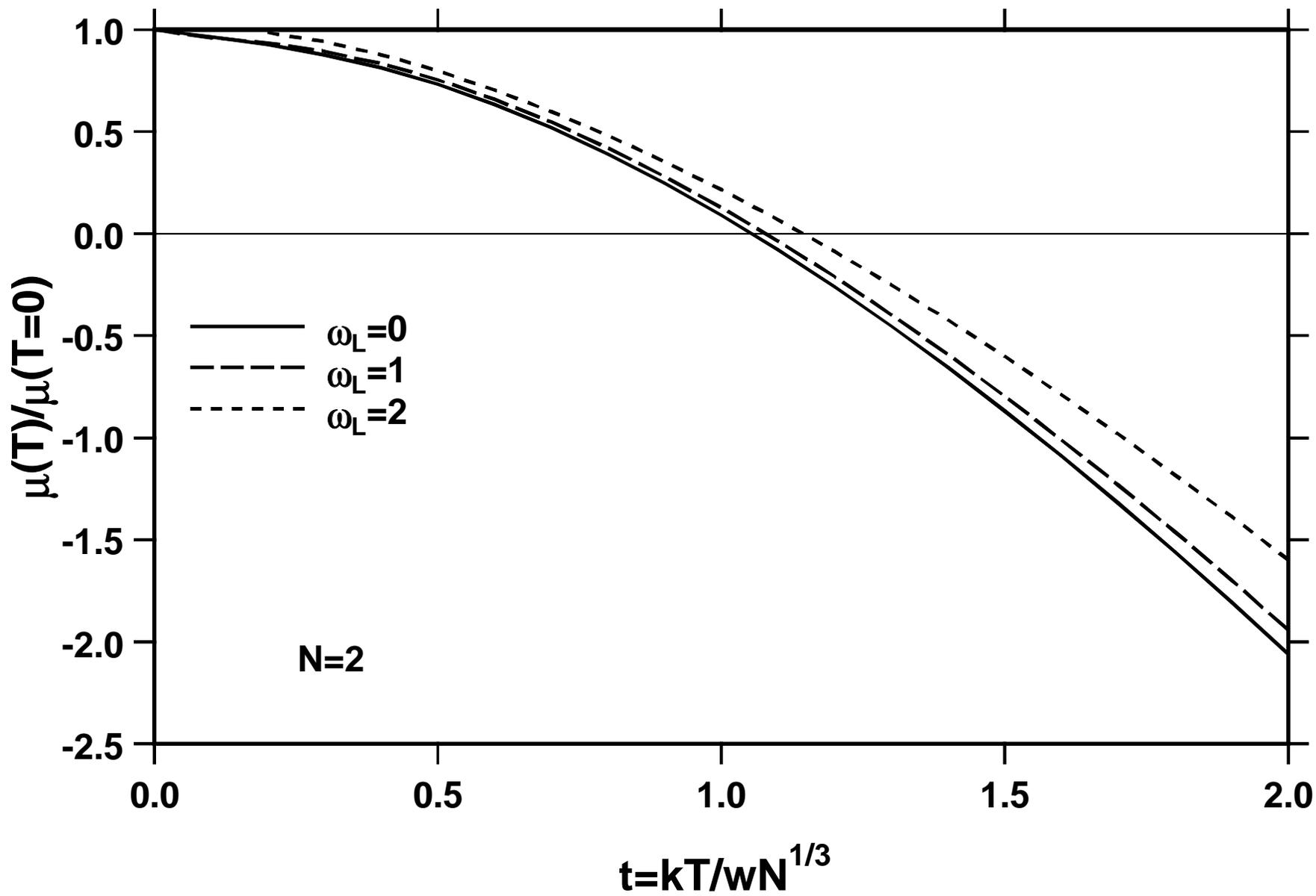
\item  {\bf Fig. 3: }Scaled chemical potential $\mu \left( T\right) /\mu
\left( T=0\right) $ as a function of the scaled temperature $t=kT/wN^{1/3}$
for 2 fermions and for $\omega _{L}/w=0$, 1 and 2.

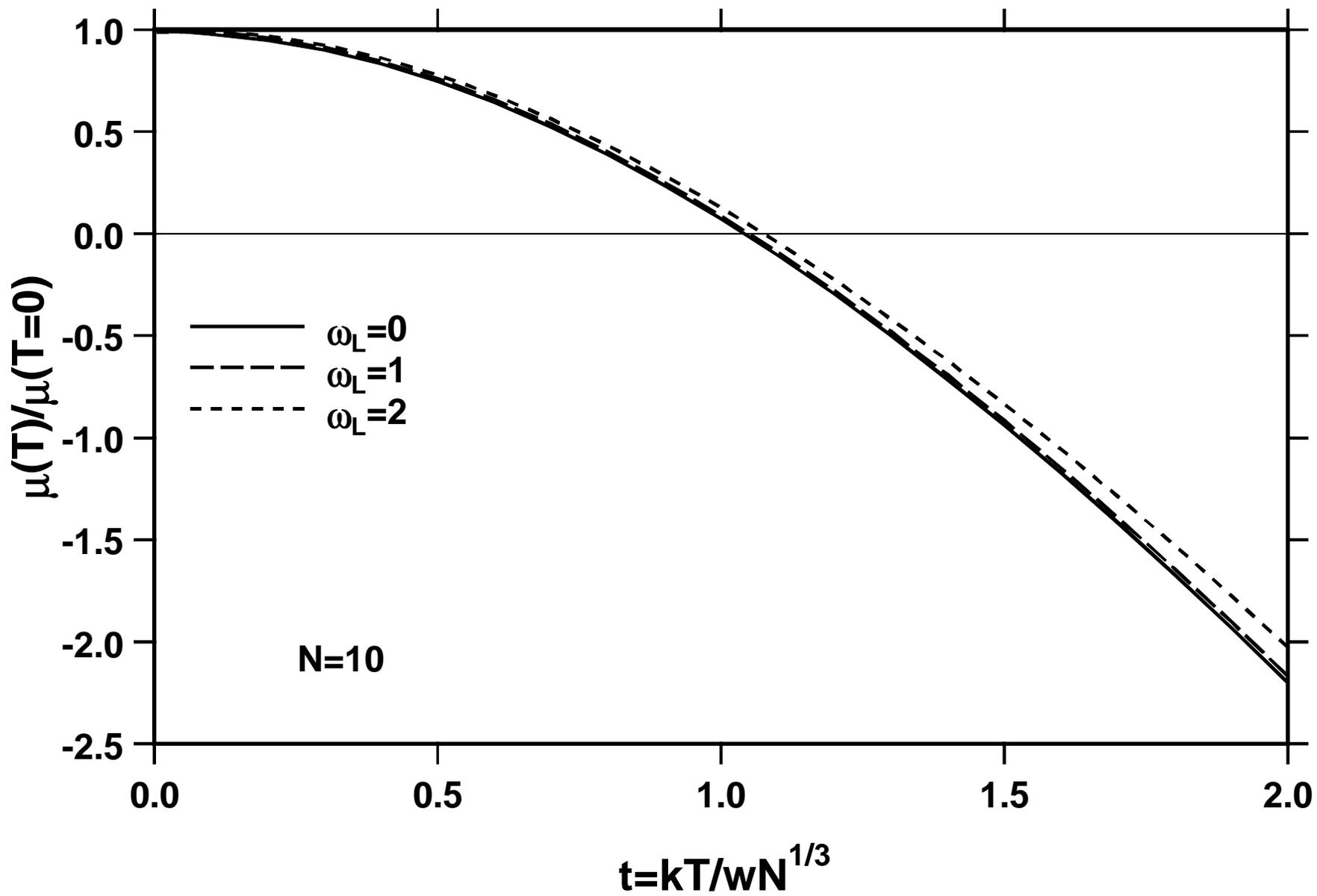
\item  {\bf Fig. 4: }Same as Fig. 4, but for 10 fermions.

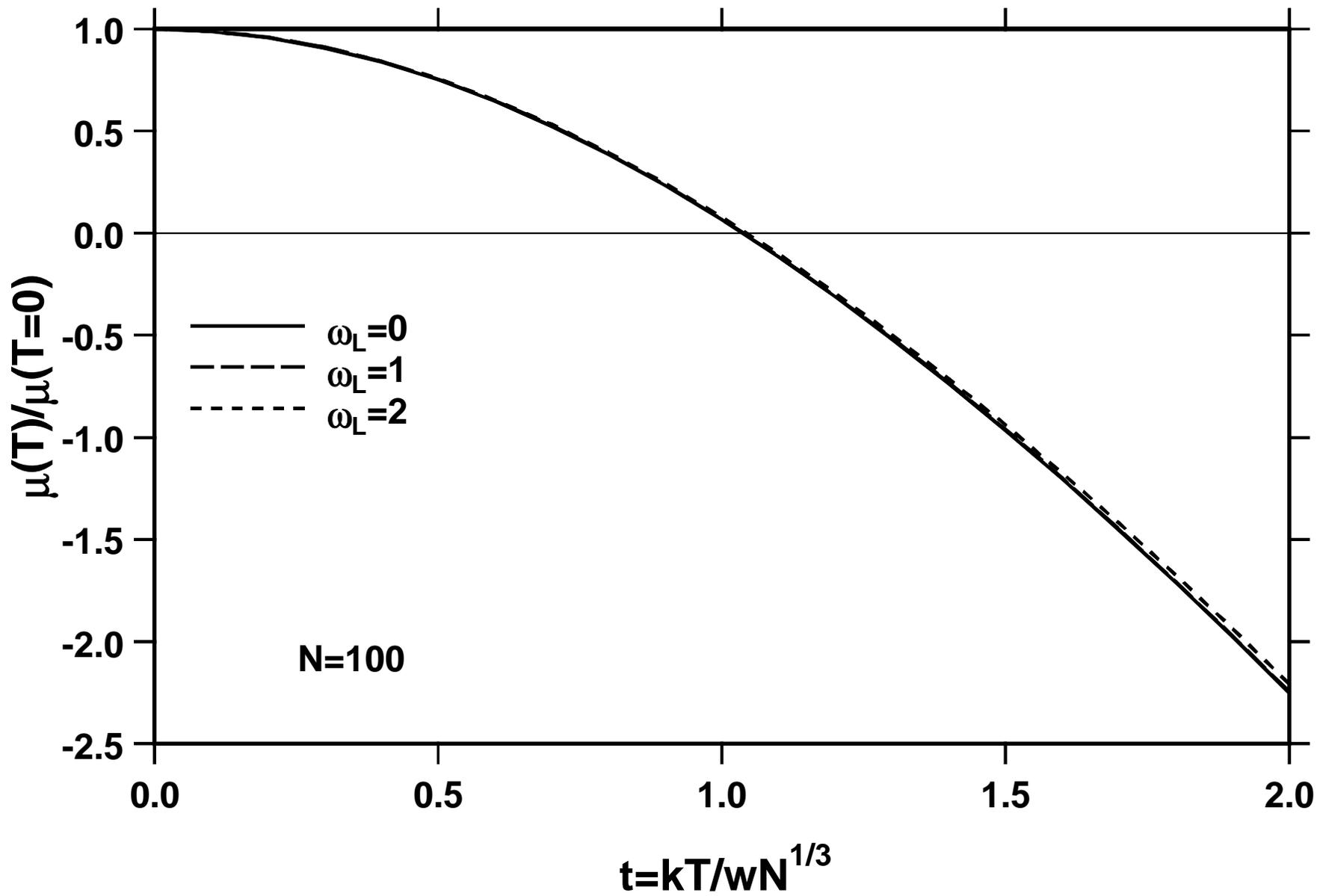
\item  {\bf Fig. 5:} Same as Fig. 4, but for 100 fermions.

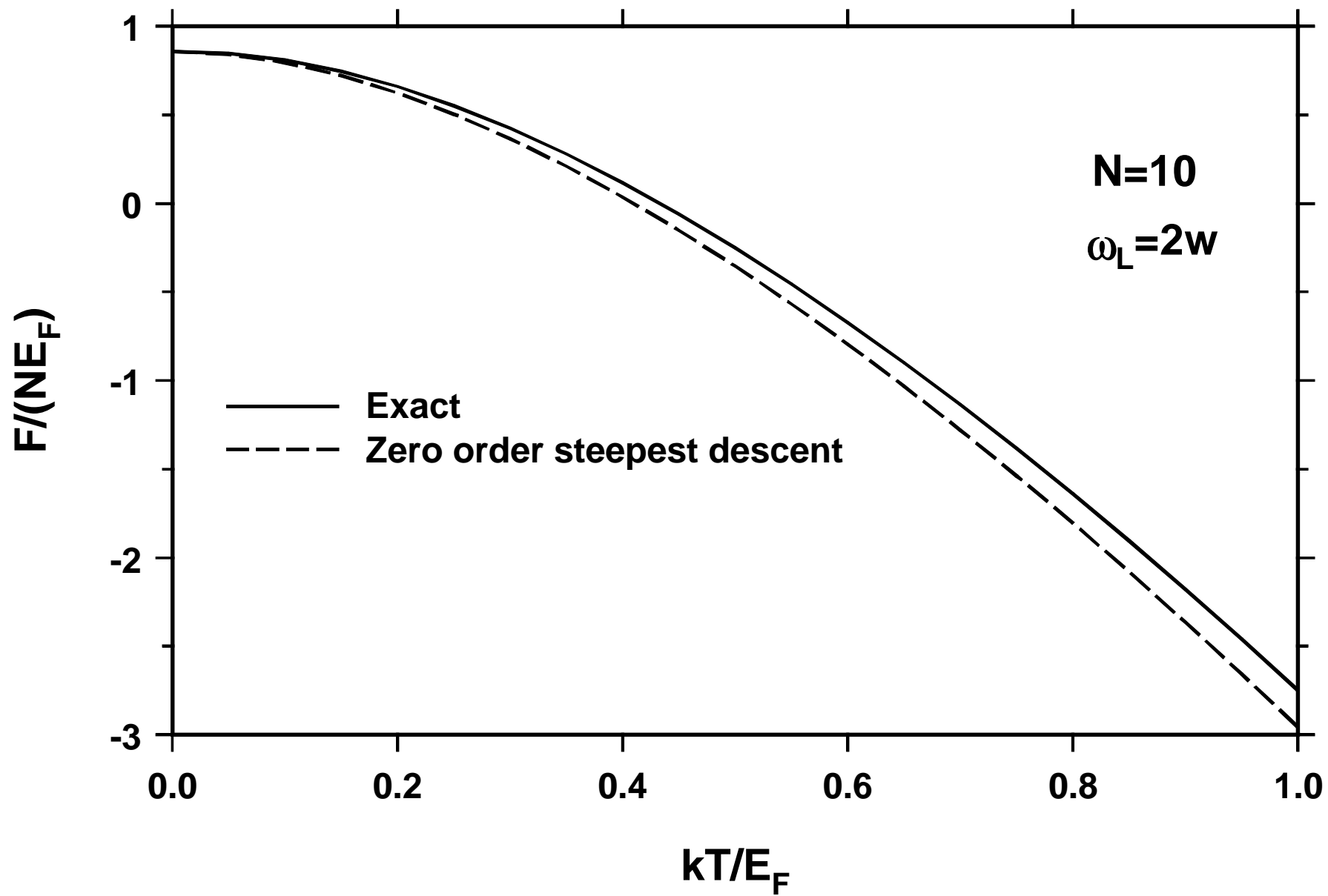
\item  {\bf Fig. 6:} Scaled free energy per particle $f={\Bbb F}_{F}/NE_{F}$
as a function of the scaled temperature $kT/E_{F}$ for 10 fermions and with
the Larmor frequency $\omega _{L}=2w$. The zero-order ``steepest descent''
contribution is indicated by the dashed line.

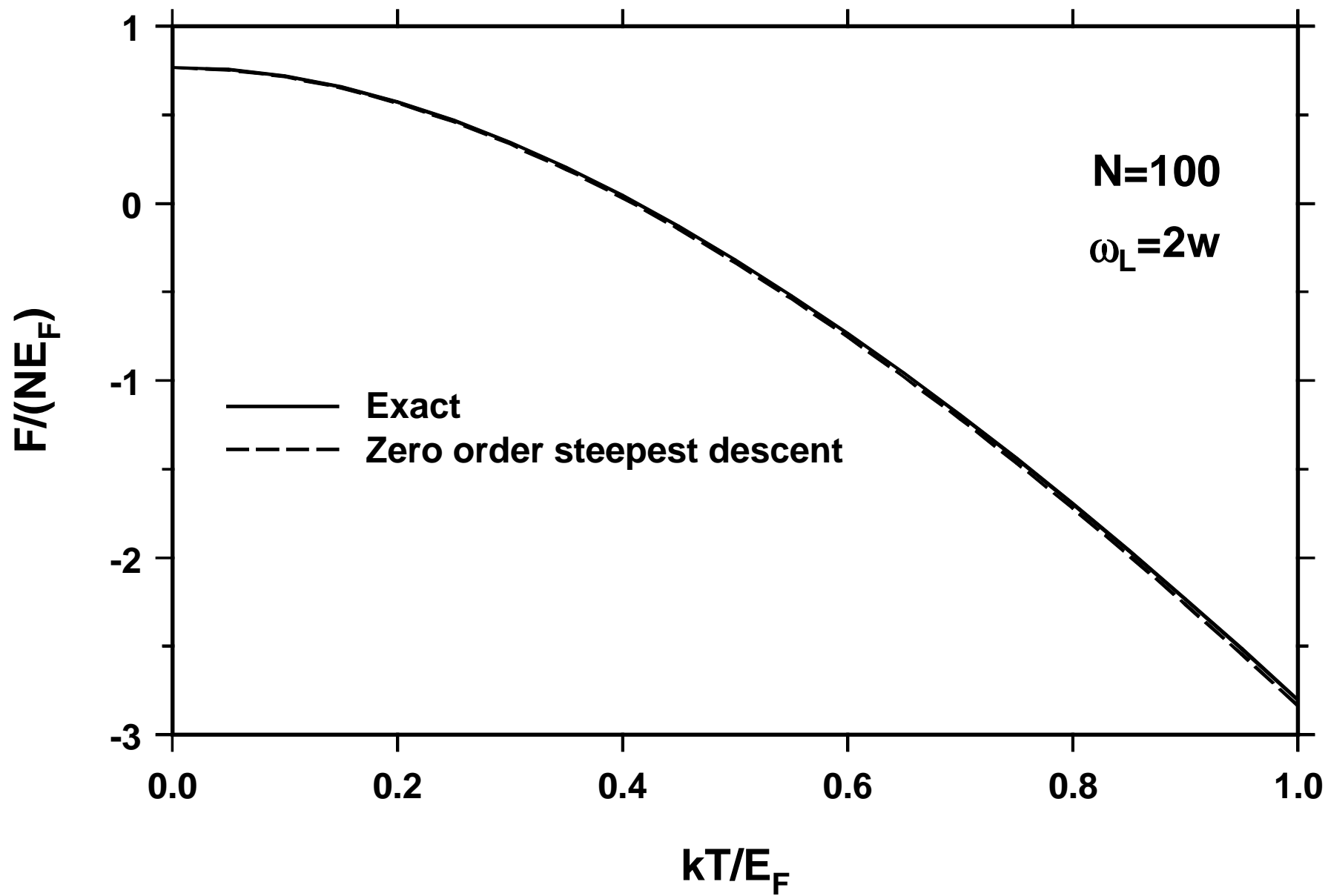
\item  {\bf Fig. 7:} Same as Fig. 6, but for 100 fermions.

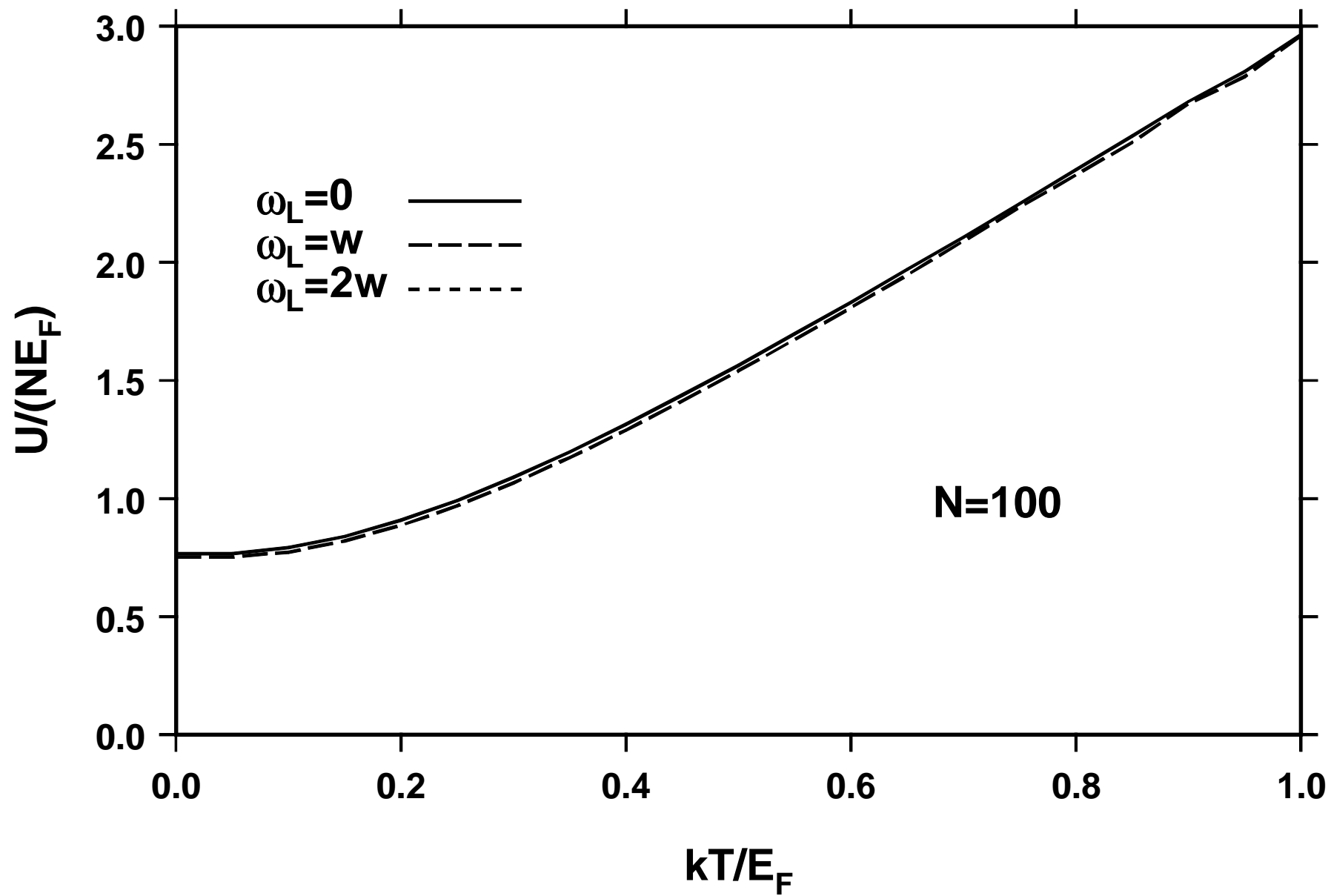
\item  {\bf Fig. 8:} Scaled internal energy per particle $u={\Bbb U}%
_{F}/NE_{F}$ for 100 fermions as a function of the scaled temperature $%
kT/E_{F}$ for several values of the Larmor frequency $\omega _{L}=0$, $w$
and $2w$.
\end{description}

\end{document}